# A Conditional Value-at-Risk Based Planning Model for Integrated Energy System with Energy Storage and Renewables

Ang Xuan[a], Xinwei Shen[a,*], Qinglai Guo[a,b] and Hongbin Sun[a,b,*]

[a]Tsinghua-Berkeley Shenzhen Institute, Tsinghua Shenzhen International Graduate School, Tsinghua University, Shenzhen 518055, China
[b]Department of Electrical Engineering, Tsinghua University, Beijing 100084, China

*Abstract*

**Owing to the potential higher energy supply efficiency and operation flexibility, integrated energy system (IES), which usually includes electric power, gas and heating/cooling systems, is considered as one of the primary forms of energy carrier in the future. However, with the increasing complexity of multiple energy devices and systems integration, IES planning is facing a significant challenge in terms of risk assessment. To this end, an energy hub (EH) planning model considering renewable energy sources (RES) and energy storage system (ESS) integration is proposed in this paper, in which the risk is measured by Conditional Value-at-Risk (CVaR). The proposed IES planning model includes two stages: 1) investment planning on equipment types and capacity (e.g., energy converters, distributed RES and ESS) and 2) optimizing the potential risk loss in operation scenarios along with confidence level and risk preference. The problem solving is accelerated by Benders Decomposition and Improved Backward Scenario Reduction Method. The numerical results illustrate the proposed method's effectiveness in balancing the potential operation risk and investment cost. Moreover, the effectiveness of reducing potential operation risk by introducing ESS and RES are also verified.**

## NOMENCLATURE

*Abbreviations:*

| | |
|---|---|
| IES | Integrated energy system |
| EH | Energy hub |
| ESS | Energy storage system |
| RES | Renewable energy sources |
| IC | Investment cost |
| OC | Operational cost |
| TC | Trading cost |
| MC | Maintenance cost |
| LC | Load shedding cost |

*Indices:*

| | |
|---|---|
| $s$ | Index of scenario |
| $t$ | Index of hour in a scenario |
| $i$ | Index of candidate device option |
| $j$ | Index of candidate capacity option |
| $m$ | Index of RES size option |
| $n$ | Index of ESS size option |
| $r$ | Index of energy type in EH |

*Parameters:*

| | |
|---|---|
| $k$ | Annualized amortization coefficient |
| $dr$ | Discount rate |
| $T$ | Lifetime |
| $p_s$ | Probability of scenario $s$ |
| $I$ | Investment cost |
| $\alpha$ | Confidence level |
| $\beta$ | Risk parameter |
| $pr^g$ | Input natural gas unit price |
| $pr_t^e$ | Input electricity unit price at hour $t$ |
| $\lambda$ | Maintenance unit cost |
| $\mu$ | Shedding load unit cost |
| $v_{in}/v_r/v_{out}$ | Cut-in/Rated/Cut-out wind speed |
| $\overline{E}$ | Maximum capacity of one ESS module |
| $\eta^{ch}/\eta^{dis}$ | Charging/Discharging efficiency of ESS |
| $C$ | Coupling coefficient in EH |
| $\overline{P_{s,t}^{out}}$ | Maximum output power |
| $L$ | Load demand |

*Variables:*

| | |
|---|---|
| $u^{i,j}$ | Binary decision variable of capacity $j$ of device $i$ |
| $z^{m/n}$ | Integer decision variable of device $m/n$ |
| $P$ | Operating power |
| $Q^g$ | Total gas consumption |
| $Q^i$ | Total output work of device $i$ |
| $Q^m$ | Total charging and discharging mileage of ESS $m$ |
| $P^{ch}/P^{dis}$ | Charging/Discharging power of ESS |
| $L^{SHED}$ | Shedding power |
| $\zeta_\alpha$ | Auxiliary variable |

The short version of this paper was presented at virtual CUE2020, OCT 10-17, 2020. This paper is a substantial extension of the short version of conference paper.
This work is supported by the National Key R&D Program of China (2018YFB0905000), the National Natural Science Foundation of China (NSFC) (52007123), and the Science, Technology and Innovation Commission of Shenzhen Municipality (No. JCYJ20170411152331932).
Corresponding: sxw.tbsi@sz.tsinghua.edu.cn (Xinwei Shen), shb@tsinghua.edu.cn (Hongbin Sun).

| | | | |
|---|---|---|---|
| $L^{'}$ | Output power of EH | $E^m$ | State of charge of ESS $m$ |
| $\overline{P^{ch}}/\overline{P^{dis}}$ | Maximum charging/discharging power | $v^{ch}/v^{dis}$ | Charging/Discharging state of ESS |

## I. INTRODUCTION

### A. Background and Motivation

ALONG with the increasing pressure on energy crisis and environmental pollution, the integrated energy system (IES) has attracted broad interests as different energy systems can be combined to achieve a higher energy supply efficiency and flexibility [1]. Energy hub (EH) [2] concept is introduced as a tool to model IES in the project, "Vision of Future Energy Networks." An EH is a group of energy facilities where the production, conversion, storage, and consumption of different energy carriers occurs, which is a promising option for IES planning. Energy storage systems (ESS) are vital in alleviating renewable energy and load fluctuations, which can provide other services, including peak shaving, uninterruptible power supply, and energy arbitrage [3]. Moreover, it is predicted that the investment and operation costs of ESSs will become more affordable [4], which has also been proved by price data from the vendors. Correspondingly, renewable energy sources (RES) can perform an essential role by addressing fossil fuel depletion and global warming, which are critically important due to their environmental-friendly nature [5].

How to efficiently assess and reduce the potential risk when planning and operating an IES is an important and essential issue due to the complexity of IES components. To tackle this problem, extensive efforts have been made to study the optimal operation of IES considering risk. However, few works have been focused on IES planning, in which the models and methods mainly focused on the co-optimization for investment and operation strategy instead of risk management.

Besides, current research has no unified standard for IES's risk management, and the definition of risk indices is relatively rough, which is primarily based on traditional power-supply reliability indices. Therefore, further research on the risk management of IES should be performed.

### B. Literature Review

Considering the differences between diverse energy systems, IES's operation and planning still faces many difficulties and needs to be further studied. Currently, several scholars have carried out related research on the operation and planning of IES. On IES operation, reference [6] proposed a mixed-integer linear programming (MILP) short-term operation model that couples power and gas networks for more flexibility and reliability. Literature [6] designed a paradigm and its operation model for interconnected EHs. Alternating direction method of multipliers (ADMM) was applied for synergistic operation of distributed IES in [8]. A coordinated regional-district operation method for IES to enhance the resilience in extreme conditions was proposed in [9]. On IES planning, an EH expansion planning model was proposed in [10], and [11] proposed a multi-stage active distribution network planning model integrated with ESS. Combining both the long-term investment and the short-term operation strategies, reference [12] presented a two-stage optimization method for a coupled IES capacity planning problem, considering economical operation and environmental issues of regional IES. In [13], an IES planning model containing an operational module that developed a steady-state optimal multi-energy flow is proposed, in which the multi-stage expansion module can also optimize the investment decisions. Meanwhile, how to decide the component capacities and operate the IES in applications were investigated in [14]-[15].

Moreover, with growing complexity, IES is facing new challenges regarding the uncertainties from different sources, e. g. RES, energy price, load demand. Stochastic optimization and robust optimization are typical methods to handle inherent uncertainties. The proposed approach in [16] adopted robust optimization to plan transmission lines to characterize the uncertainty sources pertaining to load demands and wind power productions through polyhedral uncertainty sets. To address the uncertainties of wind power output, a two-stage robust optimization model was proposed in [17] to coordinate the reactive power compensators and find a robust optimal solution. Similarly, an operation model decomposed into two subproblems representing the feasibility (security) and the optimality (economic) can be found in [18]. As for stochastic optimization, [19] presented a two-stage model for distributed energy system planning, in which Monte Carlo simulation (MCS) is applied to model energy demand and supply uncertainties. Similarly, a two-stage stochastic MILP model was proposed in [20] to invest renewable generation economically.

Conditional Value-at-Risk (CVaR) is a widely-used concept for quantifying the uncertainties and risk of portfolios in financial industry, which has been applied in power system by some scholars. It has been proved to be effective for application to maximize the profits of aggregators in [21]. It can also be applied for energy storage operation in transmission system [22] and microgrid [23], as well as quantifying risk of wind power ramps [24]. Moreover, CVaR has also been applied in economic dispatch problems in different scales. For instance, it was used to minimize the total procurement cost of conventional generation and reserve in [25]; and in [26], it was used to minimize the risk value of energy cost in day-ahead home energy management, while [27] presented a two-stage stochastic unit commitment model in transmission systems.

To conclude, we summarized the features of previous work as follows: 1) current analytical methods for addressing uncertainties in electric power system and IES such as stochastic/robust optimization [16]-[20] have been relatively mature, while the risk assessment and management considering uncertainties are somewhat roughly. 2) the applications of CVaR [21]-[27] in power systems mainly focus on operation towards one or more uncertainties such as load variations, RES, etc. 3) CVaR has been applied mostly in power system, while study on CVaR-based IES planning is still a blank.

*C. Problem Identification and Main Contributions*

Based on the literature mentioned above and the essential needs in IES applications, in this paper, CVaR approach is applied to model the risk caused by uncertainties in IES operation, which could in turn benefit the IES planning by investment risk management.

Major contributions of the paper are therefore twofold:
1) A two-stage investment-operation IES planning method is proposed based on CVaR, which can measure operation risk brought by renewable generation and load variations effectively and make a trade-off between risk and cost. With proposed model and parameters on confidence level and risk preference, the decision-makers can choose the investment strategies properly. Moreover, ESSs are included in the proposed model to improve IES's operation flexibility and proved to be cost-effective.
2) To address the computation efficiency issue brought by massive operation scenarios integrations in proposed IES planning model, we further applied an improved Backward Reduction Method in scenario reduction, cases showed the results' deviation compared with standard case and its advantages over K-means clustering. Furthermore, Benders Decomposition is also applied in the solving process to help accelerate the convergence, which outperforms the state-of-the-art commercial solver Gurobi/Cplex.

The remaining parts are structured as follows: Section II presents CVaR to quantify potential risk loss, the mathematical formulation of the planning model and the solution strategy are shown in Section III. The numerical case study and analysis are performed in section IV. Conclusions are given in section V.

## II. PRELIMINARIES OF CONDITIONAL VALUE AT RISK

Value-at-risk (*VaR*) is a measure of the risk of loss for investments. It estimates how much a set of investment portfolios might lose with a given probability, given market condition, in a set period [28]. *VaR* was firstly proposed by J. P. Morgan in the early 1990s and has become one of the most popular methods in financial institutions to measure investment portfolios' risk. Its mathematical definition is as follows:

$$\Psi(x,\varsigma) = P\{y \mid f(x,y) \leq \varsigma\} \quad (1)$$

$$VaR_\alpha = \min\{\varsigma \mid \Psi(x,\varsigma) \geq \alpha\} \quad (2)$$

In (1) $x$ is the investment decision vector representing a portfolio, and the vector $y$ stands for uncertainties governed by a probability measure $P$. For a certain $x$, we use $\Psi(x,\cdot)$ to denote the accumulative distribution function (1) for the loss $f(x,y)$. (2) is the general definition of *VaR*, $\alpha$ is the confidence level, which in some applications its value would usually be close to 1, e.g., $\alpha = 0.95$. The minimun can always be attained because $\Psi$ is continuous and increasing, $VaR_\alpha$ is the unique value satisfying $\Psi(x,\zeta) = \alpha$.

In 2008, several charges about *VaR* in Global Association of Risk Professionals Review [29] had been proposed: One was that tail risks are non-measurable; another was that *VaR* is not a coherent risk measure since it violates the sub-additivity property. Conditional Value-at-Risk (*CVaR*), also known as Mean Excess Loss, or Tail VaR, is considered to be a more consistent measure of risk than *VaR* [30]. *CVaR* is derived by taking a weighted average of the "extreme" losses in the tail of the frequency distribution function, beyond the *VaR* cut-off point in Fig.1.

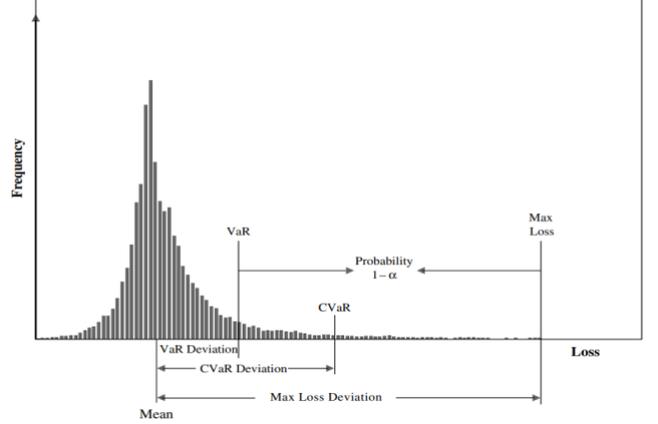

Fig.1. VaR and CVaR [32]

*CVaR* and its minimization formula were firstly proposed in [30], which demonstrated its numerical effectiveness including portfolio optimization through several case studies. *CVaR* quantifies the average loss over a specified period of unlikely scenarios beyond the confidence level. For example, a one-day 99% *CVaR* valued at $12 million means that the worst 1% scenario's expected loss over one day is $12 million. It was proved in [30] that as a function of $x$ and $VaR_\alpha$, $F(x,VaR_\alpha)$ is convex and continuously differentiable, which means $CVaR_\alpha$ associated with investment decision vector $x$, for a given confidence level $\alpha$, can be determined by the formula (3a)-(3b):

$$F(x,VaR_\alpha) = VaR_\alpha + \frac{1}{1-\alpha}\int_y [f(x,y) - VaR_\alpha]^+ p(y)dy \quad (3a)$$

where $[t]^+ = \max\{0,t\}$

$$CVaR_\alpha = \min F(x,VaR_\alpha) \quad (3b)$$

Reference [31] persisted the *CVaR* concept can be calculated in multi-scenario simulation and for general loss distributions by:

$$F(x,VaR_\alpha) = VaR_\alpha + \frac{1}{1-\alpha}\sum_{s=1}^{N} p_s [f(x,y_s) - VaR_\alpha]^+ \quad (4)$$

where $[t]^+ = \max\{0,t\}$

Due to the limitations of *VaR* in the estimation of risk and the advantages of *CVaR* compared with *VaR*, the risk management in this study is addressed by *CVaR* in (4). Therefore, for IES planning problem, considering the investment decisions $x$ as investment portfolios, $f(x,y_s)$ denotes the loss in scenario *s*, i.e. the operation cost including electricity/gas consumption......, and scenario $s = 1,\cdots,N$ have probabilities $p_s$ respectively.

## III. MODEL FORMULATION

Based on CVaR concept in previous section, a regional IES planning model is presented considering multi-energy device {combined cooling, heating, and power (CCHP), gas boiler (GB), air conditioner (AC), transformer (TX)}, RES {photovoltaics (PV), wind turbine (WT)}, and ESS {battery energy storage system(BESS), heating energy storage system (HESS), cooling energy storage system (CESS)}. It provides a novel approach for planning an EH with the balance between system's risk and economy.

### A. Two-stage Planning Model for Investment-Operation Co-optimization

As explained in the *Introduction* Section, the proposed two-stage model consists of two stages, in which the investment decisions are made in the first stage and the operation strategies are optimized in the second stage both from the system operators' perspective. As shown in Fig.2., the variables consist of binary variables $(u^{i,j})$ of multi-energy device and integer variables $(z^m, z^n)$ of RES/ESS are embedded in the first-stage problem, and the corresponding charge/discharge schedule binary decision variables $(v_{s,t}^{ch/dis,n})$ of ESSs and continuous variables $(p_{s,t}^{ch/dis,n}, p_{s,t}^{i,j}, p_{s,t}^m, L_{s,t}^{SHED,r})$ parameterize the second-stage problem. The capacity of multi-energy device, RES, as well as the capacity and maximum charging/discharging power of ESS resulting from the first-stage problem, thereby affects the decision variables evaluated in the second-stage problem. To solve the two-stage optimization model, Benders Decomposition (BD) technique [33]-[34] has been employed in many literatures.

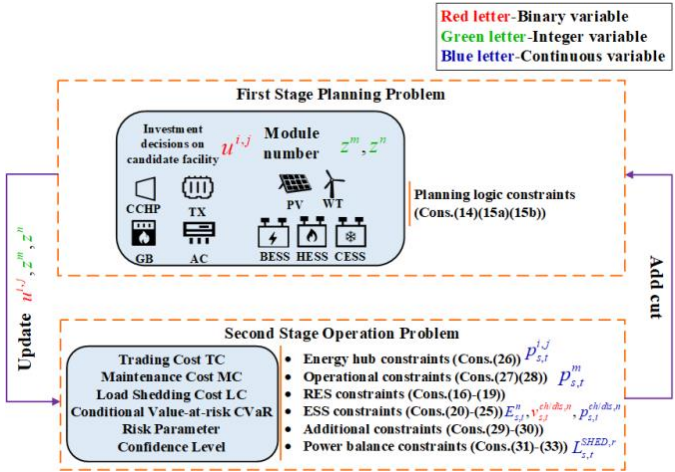

Fig.2. Structure of two-stage planning-operation optimization problem

### B. Objective Function

The objective function without *CVaR* in (5) is to minimize the total cost considering annualized investment cost and a weighted summation of expectation operation cost containing trading cost, maintenance cost, and load shedding cost in one year's multiple typical daily scenarios.

$$\min \ obj = IC + OC = IC + \sum_s p_s (TC_s + MC_s + LC_s) \quad (5)$$

The first term $IC$ in (6) of IES denotes the investment cost with annualized amortization coefficient (7) to amortize over their lifetime. We use binary variables $u^{i,j}$ to represent the selection with a specific capacity $j$ of a specific candidate device $i$, which is equal to 1 if the devices' candidate option is invested, and being 0 otherwise; meanwhile, this model includes sizing for RES/ESS, $(z^m, z^n)$ are the integer decision variables for the number of RES/ESS modules, while superscript $m$ and $n$ are device options of RES/ESS.

$$IC = \sum_i \sum_j k^{i,j} I^{i,j} u^{i,j} + \sum_m k^m I^m z^m + \sum_n k^n I^n z^n$$
$$i \in \{CCHP, GB, AC, TX\}, \quad (6)$$
$$m \in \{PV, WT\}, n \in \{BESS, TESS, CESS\}$$

$$k^\tau = \frac{dr(1+dr)^{T^\tau}}{(1+dr)^{T^\tau} - 1}, \ t \ \tau = \{i, m, n\} \quad (7)$$

In (5), $TC_s, MC_s, LC_s$ denote trading cost, maintenance cost, and load shedding cost in scenario $s$, respectively. Operation cost is the probabilistic weighted summation of scenarios, more scenarios account for a more accurate calculation result.

$$TC_s = \sum_t pr_t^e P_{s,t}^e + pr^g P_s^g \quad (8)$$

$$P_{s,t}^e = \sum_i \sum_j P_{s,t}^{i,j,e} \quad (8a)$$

$$P_s^g = \sum_t \sum_i \sum_j P_{s,t}^{i,j,g} \quad (8b)$$

$$MC_s = \sum_i \lambda^i Q_s^i + \sum_m \lambda^m Q_s^m + \sum_n \lambda^n Q_s^n \quad (9)$$

$$Q_s^i = \sum_j \sum_t P_{s,t}^{i,j} \quad (9a)$$

$$Q_s^m = \sum_t P_{s,t}^m \quad (9b)$$

$$Q_s^n = \sum_t (P_{s,t}^{ch,n} + P_{s,t}^{dis,n}) \quad (9c)$$

$$LC_s = \sum_{r \in \{e,c,h\}} \mu^r \sum_t L_{s,t}^{SHED,r} \quad (10)$$

In the superscript on the upper right corner, $g, h, c, e$ represents different energy forms: natural gas, heating, cooling, and electricity, respectively. For each scenario $s$, Trading cost $TC_s$ in (8) consists of electricity consumption $P_{s,t}^e$ in (8a) multiplied with electricity price at time $t$ $pr_t^e$ and gas consumption of all facilities, $P_s^g$ in (8b), multiplied with gas price $pr^g$. Maintenance cost $MC_s$ in (9) is related to operational power of multi-energy device, RES and charging/discharging mileage of ESS, which are denoted by (9a)-(9c), respectively. Load shedding cost (10) is calculated by multiplying unit cost of different types of loads shedding $\mu^r$ with the summation of shedding load $L_{s,t}^{SHED,r}$.

Then we introduce the Conditional Value-at-Risk (*CVaR*) of the operation cost as follows:

$$F(x,\zeta_\alpha) = \zeta_\alpha + \frac{1}{1-\alpha}\sum_{s=1}^{N}p_s[TC_s + MC_s + LC_s - \zeta_\alpha]^+ \quad (11)$$
$$CVaR_\alpha = \min F(x,\zeta_\alpha)$$

where $CVaR_\alpha$ denotes the loss expectation of the detected scenarios in (11) derived from (4) with the auxiliary variable $\zeta_\alpha$. In [30], it was proved that auxiliary variable $\zeta_\alpha$ could be approximated to $VaR_\alpha$ while $CVaR_\alpha$ reaches the optimal value. Then the final objective function considering $CVaR$ is shown as (12),

$$\min\ obj = IC + [(1-\beta)OC + \beta CVaR_\alpha] \quad (12)$$

To manage the risk in the study, the second term $[(1-\beta)OC + \beta CVaR_\alpha]$ represents the trade-off between operation cost $OC$ in all scenarios and corresponding expected risk loss. Different values of risk parameters $\beta$ between 0 and 1 are considered for the investors to address different investors' risk aversion degrees, if the value of $\beta$ is closer to 1, the significance of the risk namely CVaR, increases.

*C. Constraints*

*1) Investment logic constraints*

$$\sum_j u^{CCHP,j} \geq 1 \quad (13a)$$

$$\sum_j u^{TX,j} \geq 1 \quad (13b)$$

$$z^m \geq 0 \quad (14a)$$

$$z^n \geq 0 \quad (14b)$$

Constraints (13) and (14) model the binary decisions and integer decisions for investments on multi-energy device and RES/ESS, which means there should be at least one device of CCHP and TX constructed to complete the EH and the number of RES/ESS module should be positive.

*2) RES constraints*

$$0 \leq P_{s,t}^{WT} \leq z^{WT}\overline{P_{s,t}^{WT}} \quad (15)$$

$$\overline{P_{s,t}^{WT}} = \begin{cases} 0 & v_{s,t} \leq v_{in}, or\ v_{s,t} \geq v_{out} \\ \frac{\rho v_{s,t}^3 S^{WT}\eta^{WT}}{2} & v_{in} \leq v_{s,t} \leq v_r \\ \frac{\rho v_r^3 S^{WT}\eta^{WT}}{2} & v_r \leq v_{s,t} \leq v_{out} \end{cases} \quad (16)$$

$$0 \leq P_{s,t}^{PV} \leq z^{PV}\overline{P_{s,t}^{PV}} \quad (17)$$

$$\overline{P_{s,t}^{PV}} = H_{s,t}\cos\theta S^{PV}\eta^{MPPT}\eta^{PV} \quad (18)$$

In (16) and (18), $S^{WT}, \eta^{WT}$ imply blade area and conversion efficiency of wind turbine module, $S^{PV}, \eta^{MPPT}, \eta^{PV}$ imply PV module area, conversion efficiency of MPPT (Maximum Power Point Tracking) and PV panels, $\rho, v_{s,t}, H_{s,t}, \theta$ denote air density, wind speed, light intensity and solar tilt angle depend on weather conditions. (16) and (18) are general output power formulas of wind turbine and PV for one module, which become the outputs upper limitation through multiplying with number of modules $z^{WT}$ and $z^{PV}$ in (15) and (17).

*3) ESS constraints*

$$v_{s,t}^{ch,n} + v_{s,t}^{dis,n} \leq 1,\ v_{s,t}^{ch,n}\in\{0,1\}, v_{s,t}^{dis,n}\in\{0,1\} \quad (19)$$

$$0 \leq P_{s,t}^{ch,n} \leq M\ v_{s,t}^{ch,n} \quad (20a)$$

$$0 \leq P_{s,t}^{ch,n} \leq z^n \overline{P^{ch,n}} \quad (20b)$$

$$0 \leq P_{s,t}^{dis,n} \leq M\ v_{s,t}^{dis,n} \quad (21a)$$

$$0 \leq P_{s,t}^{dis,n} \leq z^n \overline{P^{dis,n}} \quad (21b)$$

$$0 \leq E_{s,t}^n \leq z^n \overline{E^n} \quad (22)$$

$$E_{s,t+1}^n = E_{s,t}^n + (P_{s,t}^{ch,n}\eta^{ch,n} - P_{s,t}^{dis,n}/\eta^{dis,n})\Delta t \quad (23)$$

$$E_{s,0}^n = E_{s,24}^n \quad (24)$$

Constraint (19) ensures that the ESSs cannot be charged and discharged at the same time. Charging/discharging power of ESSs are limited by the power of investment option in (20)-(21), in which the upper bound is denoted by integer variable $z^n$, the number of ESSs module and binary variable $v_{s,t}^{ch,n}/v_{s,t}^{dis,n}$ denoting the charge/discharge states, while M is a large number used in Big-M method. Constraint (22) represents the SOC (state of charge) is limited by the number of ESSs module and energy capacity for one module. Constraint (23) denote the relationship between charge/discharge power and SOC (state of charge). Constraint (24) ensures the initial and final values of SOC in a scenario are the same.

*4) Energy hub constraints*

$$\begin{bmatrix} L_{s,t}^{e'} \\ L_{s,t}^{h'} \\ L_{s,t}^{c'} \end{bmatrix} = \sum_i\sum_j \begin{bmatrix} C^{i,j,ee} & C^{i,j,eg} \\ C^{i,j,he} & C^{i,j,hg} \\ C^{i,j,ce} & C^{i,j,cg} \end{bmatrix}\begin{bmatrix} P_{s,t}^{i,j,e} \\ P_{s,t}^{i,j,g} \end{bmatrix} \quad (25)$$

$$0 \leq P_{s,t}^{i,j,e} \leq u^{i,j}\overline{P^{i,j,e}} \quad (26)$$

$$0 \leq P_{s,t}^{i,j,g} \leq u^{i,j}\overline{P^{i,j,g}} \quad (27)$$

According to the EH theorem [2], the energy coupling matrix equations is constructed as shown in (25). Constraints (26) and (27) limit the power of multi-energy coupling device with binary variable $u^{i,j}$.

*5) Additional constraints*

$$0 \leq L_{s,t}^{SHED,r} \leq L_{s,t}^r \quad (28)$$

$$0 \leq \sum_m z^m \overline{P^m} \leq \sigma(\sum_i\sum_j u^{i,j}\overline{P^{i,j,e}} + \sum_m z^m\overline{P^m}) \quad (29)$$

Constraint (28) represents the shedding load should be less than the total load. Constraint (29) limits the planned RES energy capacity maximum to a certain percentage $\sigma$ of total energy capacity in IES to avoid system instability brought by renewable power intermittence. The value of $\sigma$ is determined by system operators based on experiences and case conditions.

*6) Power balance constraints*

$$L_{s,t}^e + P_{s,t}^{dis,BESS} = L_{s,t}^{e'} + P_{s,t}^{ch,BESS} + L_{s,t}^{SHED,e} + P_{s,t}^m \quad (30)$$

$$L_{s,t}^h + P_{s,t}^{dis,TESS} \leq L_{s,t}^{h'} + P_{s,t}^{ch,TESS} + L_{s,t}^{SHED,h} \quad (31)$$

$$L_{s,t}^c + P_{s,t}^{dis,CESS} \leq L_{s,t}^{c'} + P_{s,t}^{ch,CESS} + L_{s,t}^{SHED,c} \quad (32)$$

Constraints (30)-(32) are the power balance between demand and supply. It should be noted that, the supply and demand of electric load should be strictly equal, while the heating and cooling load can be relaxed based on real-world experiences.

### D. Solution Strategy

There are 106002 continous variables and 14425 integer (14420 binary) variables in 100 scenarios, massive numbers of binary variables aggregated to model due to charge/discharge variables $\left(v_{s,t}^{ch/dis,n}\right)$ of ESSs. It can be inferred that as the number of typical daily scenarios increases, the computational burden will become extremely heavy. Therefore, it is necessary to apply some methods to accelerate computational speed while ensuring accuracy.

The following methods are considerable for reducing computational time consumption:

*1)* Apply some more efficient computing method for solving the model, like Benders' Decomposition;

*2)* Increase the convergence index, e.g., change gap tolerance from 0.01% to 0.1%;

*3)* Increase the time step in operation stage, e. g. from 1 hour to 2 hours or 4 hours.

As the number of scenarios increases, the computational burden will become extremely heavy. Therefore, it is necessary to reduce the scenario number to accelerate computational speed while ensuring accuracy. Scenario reduction techniques aggregate similar scenarios based on a particular index, currently a number of scenario reduction techniques have been proposed to make practical planning problem with massive scenarios solvable. In this work, we utilize an improved backward scenario reduction method [35], and compare the clustering results with traditional k-means clustering method [36].

The process of Improved Backward Scenario Reduction method is shown as follows:

**Step 1:** suppose D is the initial set of scenarios, J is the detected set of scenarios. Compute Kantorovich Distance (KD) for each pair of scenarios in D to form the Kantorovich Distance Matrix (KDM), KD for $\varepsilon_i$ and $\varepsilon_j$ is defined as:

$$KD(\varepsilon_i, \varepsilon_j) = \|\varepsilon_i - \varepsilon_j\|_2 \quad i\ i,j = 1,2,\cdots,N_s$$

**Step 2:** for each scenario $\varepsilon_i$, find the least value $\min\{KD(\varepsilon_i, \varepsilon_j)\}$ with nearest scenario $\varepsilon_k$ and mark it;

**Step 3:** compute the Probabilistic Distance (*PD*) for each scenario $\varepsilon_i$ corresponding $\min\{KD(\varepsilon_i, \varepsilon_j)\}$ in **Step 2**, $p_{\varepsilon_i}$ is the probability of $\varepsilon_i$;

$$PD(\varepsilon_i, \varepsilon_j) = p_{\varepsilon_i} \min\{KD(\varepsilon_i, \varepsilon_j)\} \quad i\ i,j = 1,2,\cdots,N_s$$

**Step 4:** find the least value $\min\{PD\}$ among all scenarios in D and corresponding scenario $\varepsilon_l$, add $\varepsilon_l$ to J and delete it in D;

**Step 5:** find the marked nearest scenario of $\varepsilon_l$ in **Step 2**, add the probability of $\varepsilon_l$ to the marked nearest scenario;

**Step 6:** repeat **Step 3-Step 5** until the scenario reduction requirements are met.

## IV. CASE STUDIES

### A. Basic Settings

The proposed model is applied to an EH in Fig. 3. to simulate an industrial park with electricity, heating and cooling load as well as electricity and natural gas input. Dotted lines mean they are investment options to be planned and connected. The conditions for planning, including load demand, energy price, candidate device parameter, ESS/RES information, are stated as follows [37].

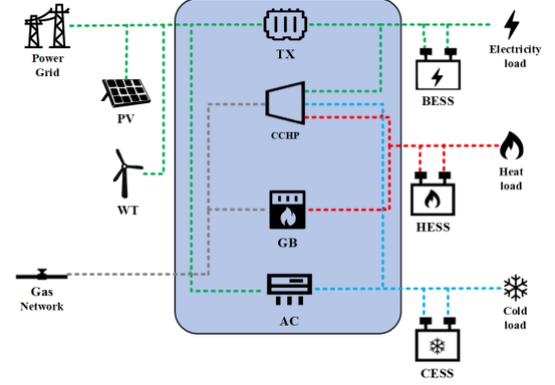

Fig.3. Energy hub model studied in the case

*1) Load Demand and Weather Condition:* The load demand is characterized by three types of load, i. e. electricity, cooling, and heating, based on an industrial park to be planned in Hebei Province, China. with 8760 hours' data in a year. The annual wind speed and light intensity are obtained from China Meteorological Data Service Centre at http://data.cma.cn/.

*2) Energy Price:* The gas price is 3.4 RMB/m3 and is considered constant in planning period. The hourly electricity price adopts peak-valley electricity price issued by the local government. No heating/cooling power is consumed at the input ports of the EH.

*3) Candidate Device Parameters:* The candidate options for energy-supply facilities are listed in Table I, with some detailed parameters illustrated in [37] due to space limit. Each multi-energy device has five types from which to choose. The maximum charging/discharging power of ESS is set to 50% of the planned capacity.

*4) RES Information:* The PV and WT candidate module parameters are listed in Table II.

The proposed model is a mixed-integer programming problem which can be solved by commercial solver such as Gurobi and Cplex. The numerical experiments of all cases are performed on a personal computer with AMD Ryzen CPU (2.10 GHz) and 20.00 GB RAM. YALMIP toolbox in MATLAB R2020b is used for modelling and Cplex 12.9.0 and Gurobi 9.0.0 optimizer for solving.

Table I
MULTI-ENERGY DEVICE AND ESS CANDIDATE PARAMETERS

|  | Candidate options | Investment cost (10$^4$RMB/MW) | Maintenance cost (RMB/MWh) |
|---|---|---|---|
| CCHP | 5 | 900 | 1.5 |
| GB | 5 | 80 | 10 |
| AC | 5 | 150 | 2 |
| TX | 5 | 30 | 10 |
| BESS | 1 MWh per module | 90 | 90 |
| HESS | 1 MWh per module | 9 | 9 |
| CESS | 1 MWh per module | 19 | 19 |

Table II
RES CANDIDATE PARAMETERS

|  | WT Module | PV Module |
|---|---|---|
| Investment cost (10⁴RMB/MW) | 350 | 600 |
| Maintenance cost (RMB/MWh) | 120 | 625 |
| Conversion Efficiency | 0.3 | 0.2 |
| Other parameters | $V_{in}$(m/s):2.5<br>$V_r$(m/s):12<br>$V_{out}$(m/s):25<br>blade length (m):40<br>air density: 1.29kg/m³ | MPPT<br>conversion efficiency:0.97<br>solar inclination:38° |

Table III
COMPARISON OF DIFFERENT RISK PREFERENCE

| Scenario=100, $\alpha$=0.95 |  | $\beta$=0.1 | $\beta$=0.5 | $\beta$=0.9 |
|---|---|---|---|---|
| Planned results | CCHP | 5MW | 5MW | 5MW |
|  | GB | 20MW | 20MW | 25MW |
|  | AC | 31MW | 36MW | 36MW |
|  | TX | 20MW | 20MW | 20MW |
|  | BESS | 15MWh | 14MWh | 14MWh |
|  | HESS | 16MWh | 42MWh | 13MWh |
|  | CESS | 13MWh | 19MWh | 21MWh |
|  | WT | 6MW | 6MW | 6MW |
|  | PV | 4120m² | 4120m² | 4120m² |
| Investment cost (×10⁴RMB) | | 25813.99 | 27870.57 | 28584.03 |
| Trading cost (×10⁴RMB) | | 14558.26 | 14578.45 | 14569.71 |
| Maintenance cost (×10⁴RMB) | | 165.13 | 165.56 | 165.54 |
| Load shedding cost (×104RMB) | | 545.13 | 91.18 | 33.04 |
| VaR(×10⁴RMB) | | 18142.76 | 18134.64 | 18104.16 |
| CVaR(×10⁴RMB) | | 28530.08 | 20013.27 | 18898.59 |
| Total cost (×10⁴RMB) | | 42408.67 | 45294.80 | 47069.59 |

*B. Analysis of investment strategies*

To address different investors' risk preferences, different values of risk parameters are considered. In general, we describe investors with risk parameter $\beta > 0.5$ as "*risk-averse investors*", investors with risk parameter $\beta < 0.5$ as "*risk-seeking investors*", investors with risk parameter $\beta$=0.5 as "*risk-neutral investors*". In the objective function (12), the decision variables of investment terms are $\{u^{i,j}, z^m, z^n\}$, which limit the operation bounds of devices. We explored the investment costs variation of varying confidence levels and risk parameters to analyze the investment strategy in Fig.4.. The value of the risk parameter $\beta$ represents the degree of investor aversion to risk, and the confidence level $\alpha$ describes the probability threshold of the expected loss.

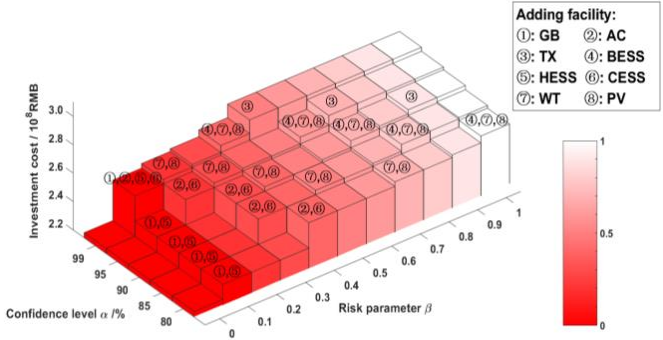

Fig. 4. Investment strategies and changes in different confidence level and risk parameter

In Fig.4., we observed that: 1) With higher confidence level ($\alpha$) and risk parameter ($\beta$) requirements, more investments on energy-supply facilities are tending to be made. 2) Multi-energy devices and ESSs often occur in additional investments simultaneously, such as GB&HESS (①&⑤), AC&CESS(②&⑥), WT(PV)&BESS(④⑦&⑧). 3) At the same confidence level, *risk-seeking investors* prefer heating and cooling facilities as additional investments due to their relatively low investment costs; as for *risk-averse investors*, they add electric facilities based on *risk-seeking investors*' investment strategies. 4) At higher confidence levels, similar investment strategies could be made in advance along with increasing risk parameters. Overall, Fig.4. can be used as an auxiliary tool to provide customized investment advice for different investors, and this is also the novelty of the model.

The planning results considering different risk preference are also shown in Table III, shedding load unit cost $\mu^e, \mu^h, \mu^c$ are set to 2, 1.8, 1.8RMB/kWh. With risk parameter $\beta$ increases, although the investment cost has increased by 7.97%, 10.73%, the trading cost and maintenance cost are still similar, while the load shedding cost has been reduced by 83.27% and 93.94%, with *CVaR* at $\beta$=0.9 was also reduced to 66.24% of the original level ($\beta$=0.1). The planning results show that the load shedding issue can be addressed by increasing risk parameter $\beta$, because of load shedding cost, which is the main reason for variations of OC and *CVaR*. Moreover, energy supply facilities with larger capacity are likely to be invested with increasing risk parameter $\beta$ due to the risk concerns.

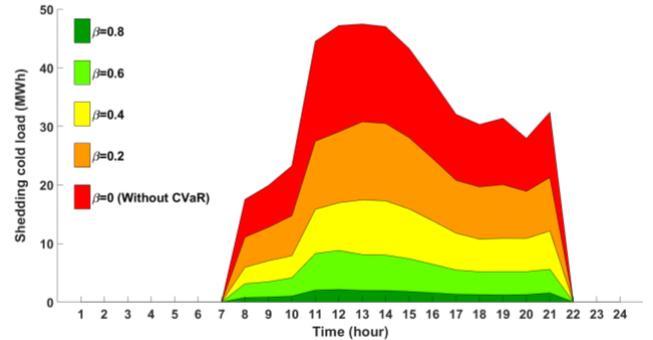

Fig. 5. Hourly shedding cold load in different risk parameter

Fig.5. takes the shedding cooling load as an example to further illustrate the superiority of the planning method in reducing shedding load considering *CVaR*. As the increasing risk aversion, the quantity of shedding cooling load decreases until it is reduced to nearly zero when $\beta \geq 0.8$, it means planning schemes will reduce adverse effects by extreme fluctuations considering load uncertainties with more investments in higher risk aversion.

*C. Analysis of different coupling conditions*

To further verify the model effectiveness, four cases considering different coupling conditions are designed to analyze the planning results in this paper:

**Case 1**-plan EH with multi-energy device (CCHP, GB, AC and TX);

**Case 2**-plan EH with multi-energy device (CCHP, GB, AC, TX) and RES (WT, PV);

**Case 3**-plan EH with multi-energy device (CCHP, GB, AC and TX) and ESS (BESS, HESS, CESS);

**Case 4**-plan EH with multi-energy device (CCHP, GB, AC and TX), RES (WT, PV) and ESS (BESS, HESS, CESS).

The specific planning results are shown in Table IV. The trading cost, maintenance cost and load shedding cost in Case 2, Case 3 and Case 4 has been reduced compared to that of Case 1. This is related to the superiority of ESS and RES in O&M cost, optimizing scheduling, and energy arbitrage compared to multi-energy coupling device. It is believed this trend will become more evident with the improvement of corresponding technology. Similar trends can be found at *CVaR* and *VaR* as well, which indicates the potential loss, or "Tail VaR" (See *Section II.*), could be reduced with RES and ESS aggregations. Results also illustrate that *CVaR* can not only provide investors with auxiliary investment strategies but also can be used as a risk measurement index to express the potential risk loss of the entire system.

Table IV
COMPARISON OF DIFFERENT COUPLING CONDITIONS

| Scenario=100, $\alpha$=0.95, $\beta$=0.5 | | **Case1** | **Case2** | **Case3** | **Case4** |
|---|---|---|---|---|---|
| Planned results | CCHP | 5MW | 5MW | 5MW | 5MW |
| | GB | 25MW | 25MW | 25MW | 20MW |
| | AC | 38.5MW | 38.5MW | 36MW | 36MW |
| | TX | 20MW | 20MW | 20MW | 20MW |
| | BESS | - | - | 1MWh | 14MWh |
| | HESS | - | - | 42MWh | 16MWh |
| | CESS | - | - | 17MWh | 8MWh |
| | WT | - | 6 | - | 42MW |
| | PV | - | 4120m$^2$ | - | 4120m$^2$ |
| Investment cost ($\times 10^4$RMB) | | 26739.25 | 27366.34 | 26458.72 | 27870.57 |
| Trading cost ($\times 10^4$RMB) | | 17948.96 | 15451.78 | 17810.78 | 14578.45 |
| Maintenance cost ($\times 10^4$RMB) | | 191.07 | 161.74 | 192.20 | 165.56 |
| Load shedding cost ($\times 10^4$RMB) | | 151.46 | 151.46 | 107.48 | 91.18 |
| VaR($\times 10^4$RMB) | | 20956.41 | 18896.09 | 20927.49 | 18134.64 |
| CVaR($\times 10^4$RMB) | | 24113.34 | 21788.36 | 23394.08 | 20013.27 |
| Total cost ($\times 10^4$RMB) | | 47941.66 | 46143.01 | 47210.98 | 45294.80 |

The hourly electricity transaction summation of all scenarios in Fig.6. can further illustrate the performance of RES and ESS in daily scheduling and the energy arbitrage with the peak-valley electricity price difference. During the valley hours (23:00-7:00), Case 3 and Case 4 are significantly higher than Case 1 and Case 2 because of BESS's charging power, thus during other periods the purchased power is lower than the above period due to discharging power. Compared with Case 1/3, the purchased electricity in Case 2/4 is always at a lower level due to local wind energy and solar energy utilization.

### D. Analysis of computational speed and accuracy

In this study, a calendar day containing 24 hours is considered to be one scenario, different scenario numbers after reduction are set from 365 scenarios in one year to relieve the computational burdens. We select 10, 30, 50, 100, 200 and 300 scenarios, respectively, and these reduced scenarios are usually called typical days. Heuristic feasible solution search is turned off in Gurobi and Cplex, with optimality gap tolerance set to 0.0001. Compared time consumptions and objective with different scenario numbers in Table V. The calculation results are the same because a certain MIP gap is set among three solution methods. In that case, we could verify the acceleration effectiveness of Benders decomposition by comparing the time consumption. The results proved the effectiveness of BD to help accelerate the convergence in solving process.

Table V
COMPARISON OF COMPUTATION TIME AND SOLUTION IN DIFFERENT SCENARIOS

| Scenario ($\alpha$=0.95, $\beta$=0.5) | Method | Time (s) | Objective ($\times 10^4$RMB) |
|---|---|---|---|
| 10 | Gurobi | 12.32 | 40769.85 |
| | Cplex | 11.26 | 40769.85 |
| | Benders | 1.98 | 40769.85 |
| 30 | Gurobi | 50.40 | 42934.94 |
| | Cplex | 65.42 | 42934.94 |
| | Benders | 11.91 | 42934.94 |
| 50 | Gurobi | 55.76 | 43791.01 |
| | Cplex | 54.12 | 43791.01 |
| | Benders | 15.25 | 43791.01 |
| 100 | Gurobi | 319.44 | 44366.26 |
| | Cplex | 287.84 | 44366.26 |
| | Benders | 49.30 | 44366.26 |
| 200 | Gurobi | 815.84 | 44927.91 |
| | Cplex | 945.60 | 44927.91 |
| | Benders | 382.92 | 44927.91 |
| 300 | Gurobi | 2022.94 | 45296.15 |
| | Cplex | 1547.94 | 45296.15 |
| | Benders | 623.13 | 45296.15 |
| 365 (Without reduction) | Gurobi | 2479.88 | 45294.80 |
| | Cplex | 2753.12 | 45294.80 |
| | Benders | 1132.61 | 45294.80 |

We choose the result based on 8760-hour datasets (i. e. 365 typical days without scenario reduction) as a benchmark to calculate the results' deviation: a positive sign means exceeding the actual value, and a negative sign means lower than the actual value. The k-means clustering, as a well-known scenario reduction method, is used to compare with the proposed improved backward reduction method in Table VI.

Table VI
COMPARISON OF CALCULATION DEVIATION

| Scenario ($\alpha$=0.95, $\beta$=0.5) | | IC | TC | MC | LC | CVaR | Total Cost |
|---|---|---|---|---|---|---|---|
| 10 | ① | -15.39% | -5.28% | +16.75% | - | -25.52% | -13.90% |
| | ② | -12.16% | -3.20% | +12.12% | - | -20.20% | -9.99% |
| 30 | ① | -11.35% | -3.77% | +5.41% | -97.92% | -21.23% | -6.74% |
| | ② | -9.08% | -2.65% | +4.34% | -82.45% | -18.34% | -5.21% |
| 50 | ① | -10.72% | -2.12% | -3.86% | -50.84% | -15.86% | -5.44% |
| | ② | -8.76% | -1.28% | -3.12% | -38.11% | -13.12% | -3.32% |
| 100 | ① | -5.72‰ | -9.25‰ | -2.84% | -15.27% | -12.46% | -3.82% |
| | ② | -4.44‰ | -2.74‰ | -1.75% | -10.21% | -9.44% | -2.05% |
| 200 | ① | -5.75‰ | +2.68‰ | -2.71% | -8.14% | -5.69% | -1.84% |
| | ② | -4.74‰ | -0.25‰ | -2.18% | -5.23% | -2.33% | -0.81% |
| 300 | ① | +0.17‰ | +0.67‰ | -2.35% | -5.78% | -2.81% | -0.05‰ |
| | ② | -0.04‰ | -0.03‰ | -7.78% | -3.90% | +9.07% | +0.03‰ |
| 365 | | 0 | 0 | 0 | 0 | 0 | 0 |

① K-means clustering method (Euclidean Distance)
② Improved backward scenario reduction method

Table VI shows that the deviations with the benchmark are smaller using proposed improved backward reduction method, compared with k-means clustering method. The proposed method considers the Probability Distance based on Euclidean distance instead of the Within-Cluster Sum of Squares (WCSS)

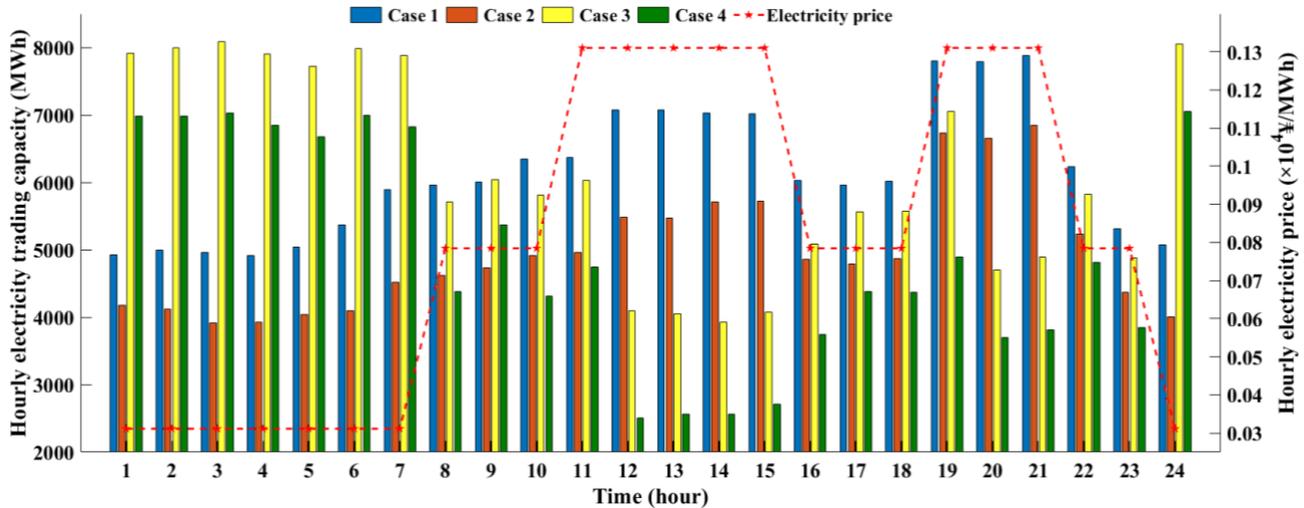

Fig. 6. Hourly electricity trading capacity under different coupling conditions

[38] in k-means clustering, while the effect of low-probability scenarios on the reduction result is preferentially excluded. Therefore, this method is more applicable for scenario sets with scattered probability distributions, it is also one of its advantages.

Moreover, the k-means method has several disadvantages: 1) the value of K needs to be set manually, and the results obtained with different K are different; 2) it's sensitive to the initial cluster center, different cluster center selection methods will produce different clustering results; 3) it's sensitive to outliers, while the proposed method overcomes these shortcomings.

## V. Conclusions

This paper presents a two-stage IES planning model considering *CVaR* aggregated with RES and ESS. The proposed model effectively co-optimizes the planning strategy and the operation strategy of IES, as well as the assessment on *CVaR*. In the proposed model, three indices on maintenance cost, trading cost, and load shedding cost are included in the objective function to optimize the annualized investment cost and operating conditions. The planning decisions, including configurations of multi-energy devices, RES and ESS module, and operation strategies for multi-energy device, RES and charging/discharging ESS, are optimized in the model. In terms of risk assessment, a CVaR based analysis is implemented in different confidence level with different risk preference.

The case studies demonstrate the effectiveness of the proposed model and illustrate the benefits of RES and ESS applications by CVaR in the IES planning and operation. ESS located on the demand side can benefit system operation by peak-valley load shifting and energy arbitrage to enhance resilience. And increasing risk parameter could benefit IES planning by eliminating load shedding caused by load and renewables' uncertainties.

Since the model is time-consuming when applied to the IES planning with hundreds of scenarios, we also propose to speed up the convergence by the improved backward reduction method and benders decomposition, which is proved to be effective, too. Further research on this topic will help balance the convergence speed and planning results' deviations.